\documentclass[twocolumn]{aastex62}

\usepackage{xspace}

\newcommand\msun {$M_{\odot}$\xspace}
\newcommand\rsun {$R_{\odot}$\xspace}
\def\flx{erg cm$^{-2}$ s$^{-1}$\xspace}
\def\lum{erg s$^{-1}$\xspace}
\def\vel{km s$^{-1}$\xspace}
\def\xmm{{\it XMM-Newton}\xspace}
\def\chan{{\it Chandra}\xspace}
\def\hst{{\it HST}\xspace}
\def\swift{{\it Swift}\xspace}
\def\nustar{{\it NuSTAR}\xspace}

\def\xshoo{X-shooter\xspace}
\def\spitzer{{\it Spitzer}\xspace}
\def\nicer{{\it NICER}\xspace}

\newcommand{\xone}{M82~X-1\xspace}
\newcommand{\xtwo}{M82~X-2\xspace}

\newcommand{\pth}{NGC\,7793~P13\xspace}
\newcommand{\snulx}{NGC~300~ULX-1\xspace}

\newcommand{\sps}{\ensuremath{\text{s\,s}^{-1}}\xspace}

\def\Teff{$T_{\rm eff}$\xspace}
\def\logg{$\log g$\xspace}

\submitjournal{ApJL}

\shorttitle{}
\shortauthors{Heida et al.}

\turnoffeditone
\begin{document}

\title{Discovery of a red supergiant donor star in SN2010da/NGC 300 ULX-1}

\correspondingauthor{Marianne Heida}
\email{mheida@caltech.edu}

\author[0000-0002-1082-7496]{M.~Heida}
\affil{Cahill Center for Astronomy and Astrophysics, California Institute of Technology, 1200 E. California Blvd, Pasadena, CA 91125, USA}

\author{R.M.~Lau}
\affil{Institute of Space \& Astronautical Science, Japan Aerospace Exploration Agency, 3-1-1 Yoshinodai, Chuo-ku, Sagamihara, Kanagawa 252-5210, Japan}

\author[0000-0002-2010-2122]{B.~Davies}
\affil{Astrophysics Research Institute, Liverpool John Moores University, Liverpool Science Park ic2, 146 Brownlow Hill, Liverpool L3 5RF, UK}


\author[0000-0002-8147-2602]{M.~Brightman}
\affil{Cahill Center for Astronomy and Astrophysics, California Institute of Technology, 1200 E. California Blvd, Pasadena, CA 91125, USA}

\author[0000-0003-0388-0560]{F.~F{\"u}rst}
\affil{Quasar Ltd for ESA, European Space Astronomy Centre (ESA/ESAC), Operations Department, 28692 Villanueva de la Ca\~nada, Madrid, Spain}

\author[0000-0002-1984-2932]{B.W.~Grefenstette}
\affil{Cahill Center for Astronomy and Astrophysics, California Institute of Technology, 1200 E. California Blvd, Pasadena, CA 91125, USA}

\author[0000-0002-6745-4790]{J.A.~Kennea}
\affil{Department of Astronomy and Astrophysics, Eberly College of Science, The Pennsylvania State University, 525 Davey Laboratory, University Park, PA 16802, USA}


\author[0000-0001-8631-7700]{F.~Tramper}
\affil{IAASARS, National Observatory of Athens, Vas. Pavlou and I. Metaxa, Penteli 15236, Greece}

\author{D.J.~Walton}
\affil{Institute of Astronomy, Madingley Road, Cambridge CB3 0HA, UK}

\author{F.A.~Harrison}
\affil{Cahill Center for Astronomy and Astrophysics, California Institute of Technology, 1200 E. California Blvd, Pasadena, CA 91125, USA}

\begin{abstract}
SN2010da/\snulx was first detected as a supernova impostor in May 2010 and was recently discovered to be a pulsating ultraluminous X-ray source. In this letter, we present VLT/\xshoo spectra of this source obtained in October 2018, covering the wavelength range 350-2300 nm. The $J$- and $H$-bands clearly show the presence of a red supergiant donor star that is best matched by a MARCS stellar atmosphere with $T_{\rm eff} = 3650 - 3900$ K and $\log(L_{\rm bol}/L_{\odot}) = 4.25\pm0.10$, which yields a stellar radius $R = 310 \pm 70 R_{\odot}$. To fit the full spectrum, two additional components are required: a blue excess that can be fitted either by a hot blackbody (T $\gtrsim 20,000$ K) or a power law (\edit1{spectral index} $\alpha \approx 4$) and is likely due to X-ray emission reprocessed in the outer accretion disk or the donor star; and a red excess that is well fitted by a blackbody with a temperature of $\sim 1100$ K, and is likely due to warm dust in the vicinity of SN2010da. The presence of a red supergiant in this system implies an orbital period of at least 0.8-2.1 years, assuming Roche lobe overflow. Given the large donor-to-compact object mass ratio, orbital modulations of the radial velocity of the red supergiant are likely undetectable. However, the radial velocity amplitude of the neutron star is large enough (up to 40-60 \vel) to potentially be measured in the future, unless the system is viewed at a very unfavorable inclination.
\end{abstract}


\section{Introduction} \label{sec:intro}

\subsection{SN2010da}
\object{SN2010da} in NGC~300 was initially discovered as a bright optical transient in May 2010 \citep{monard10}. Subsequent multi-wavelength observations indicated that the transient was probably due to an outburst of a dust-enshrouded massive star rather than a supernova \citep{chornock10, eliasrosa10, prieto10}, which put it in the diverse category of supernova `impostors' or intermediate luminosity optical transients (ILOTs). These are distinct from typical core-collapse supernovae since the impostor progenitor survives the explosion, which exhibits typical peak absolute $V$-band magnitudes ranging from -11 to -14 \citep[e.g.,][]{vandyk00,smith09}.

The progenitor of SN2010da was heavily obscured by dust; before May 2010 the system was detected in the mid-infrared with the \spitzer {\it Space Telescope} \citep{lau16}, but pre-outburst optical observations yielded only upper limits (with the exception of a single detection in the $i$-band in September 2008, at an AB magnitude of $\sim 24.2$, \citealt{villar16}). Most of the dust seems to have been destroyed in the outburst, revealing a variable optical source with $g', r', i' \approx 20-21$; post-outburst observations also revealed re-brightening mid-infrared emission \citep{lau16,villar16}. 

The massive star in the system has been suggested to be a luminous blue variable \citep[LBV,][]{binder11, binder16}, an sgB[e] star \citep{lau16}, or a yellow supergiant \citep{villar16} based on its post-outburst photometry. Optical spectra of the source are dominated by emission lines indicative of $\sim 1000$ \vel outflows and X-ray-ionized material, but no photospheric absorption lines from the star have been reported previously \citep{villar16,binder18}.

\subsection{\snulx}
In the years following the 2010 outburst, a compact, persistent X-ray source was detected at the position of SN2010da with a 0.3-10 keV luminosity of $10^{36} - 10^{37}$ \lum. This led to the suggestion that the system could be a high-mass X-ray binary (HMXB), powered by accretion onto a black hole or neutron star \citep{binder11, binder16}. As this source was never detected in X-rays before the 2010 outburst, this may have been the first observed example of a massive binary evolving into an HMXB. 

Then, in a deep \nustar+\xmm observation taken in 2016, SN2010da was detected at a much higher X-ray luminosity of $\sim 5 \times 10^{39}$ \lum \citep{carpano18}, above the threshold of $10^{39}$ \lum for ultraluminous X-ray sources (ULXs, see \citealt{kaaret17} for a recent review). Moreover, \citet{carpano18} discovered pulsations with a period of 31.6 s in its X-ray light curve, identifying SN2010da (now dubbed \snulx; not to be confused with NGC~300 X-1, a Wolf-Rayet - black hole X-ray binary in the same galaxy) as a ULX pulsar. Only a handful of ULXs have been found to harbor neutron star accretors since the first ULX pulsar was discovered in 2014 \citep{bachetti14}, although similarities between the X-ray spectra of these pulsars and the general ULX population indicate that a majority of ULXs may in fact contain neutron stars \citep{walton18}. These systems offer an opportunity to study super-Eddington accretion flows and the role of the magnetic field. \snulx is the nearest of all currently known ULX pulsars, at a distance of 2.0 Mpc \citep{2009ApJS..183...67D}, making it a valuable addition to this small sample.

Follow-up observations with \nustar, \swift and \chan in 2017 and early 2018 showed that both the pulsations and the super-Eddington luminosities were persistent during that time. Evidence for an ultra-fast outflow and possibly a cyclotron resonance feature was also found in deep \nustar and \xmm observations \citep{kosec18, walton18b}. Regular monitoring showed that the source started to fade in the summer of 2018 and was no longer detected in short \swift observations as of May 2019. \snulx was found to spin up very rapidly (at $\sim 10^{-7}$ \sps), and has one of the highest pulsed fractions seen in any ULX pulsar \citep[$\sim 55\%$ in the \xmm data and increasing towards higher energies,][]{carpano18}. The pulse period evolution thus far shows no sign of orbital modulations, ruling out an orbital period shorter than a year \citep{vasilopoulos18,ray18}.

In this letter, we present the near-UV through near-infrared (NIR) spectrum of SN2010da/\snulx obtained with VLT/\xshoo when the source was fading in X-rays. We detect photospheric absorption lines in the NIR part of the spectrum, allowing us to reliably identify the donor star as a red supergiant (RSG). We also discuss the presence of two additional components in the broadband spectrum, as well as the properties of some of the emission lines that dominate the optical part of the spectrum. 

\section{Observations and data reduction} \label{sec:obs}
We obtained VLT/\xshoo \citep{vernet11} spectra of the optical counterpart of \snulx{} \edit1{(located at RA = 00:55:04.86, Dec. = -37:41:43.7)} observed in two identical observing blocks (OBs) as part of ESO programme 0102.D-0535(A), on October 10 and October 12, 2018 (MJD 58402.1577 and 58404.2895). One of the OBs was also observed on October 9 but interrupted by a rapid response mode observation. We used the nod-on-slit mode with an ABBA nodding pattern and a nod throw of 3''. The slit widths are 0.6'' in the NIR arm, 0.7'' in the VIS arm and 0.8'' in the UVB arm, yielding nominal spectral resolutions ($\lambda/\Delta\lambda$) of 8100, 11400, and 6700, respectively.  Each OB consists of 1 ABBA sequence with individual exposure times of 600 s in the NIR and UVB arms and 511 s in the VIS arm. This provides total exposure times of 4800 s in the NIR and UVB arms and 4088 s in the VIS arm. Seeing conditions were better than 0.6'' in the first OB and better than 0.9'' in the second, meaning that the actual spectral resolution for the first OB is set by the seeing rather than the slit width and is somewhat better than nominal, while for the second OB the spectral resolution is limited by the slit width.

We reduce the data with the standard ESO pipeline for \xshoo (v3.2.0) in Reflex (v2.9.1). We use nodding mode and standard settings for all steps except the extraction, where we set {\it extract\_method} to Localization and {\it localize\_method} to Manual, with {\it localize\_slit\_hheight} at 1.5'' for all arms. We also increase {\it crsiglim} to 10 for the VIS and UVB arms and 15 for the NIR arm. 

\begin{figure*}
    \centering
    \includegraphics[width=\textwidth]{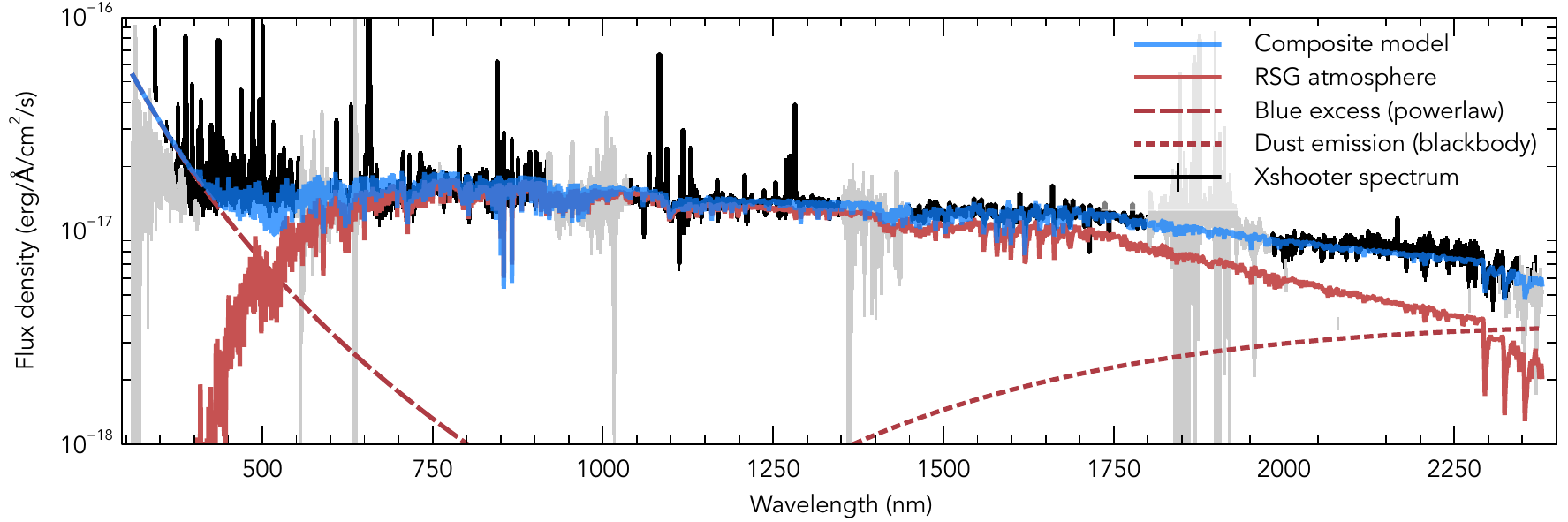}
    \caption{The full \xshoo spectrum (black, with errorbars) with areas heavily affected by telluric emission lines shown in grey. Overplotted are the best matching MARCS model atmosphere (solid red line), a 1100 K blackbody to account for excess red emission (red dotted line) and a power law with $\alpha \approx 4$ to account for excess blue emission (red dashed line, although the shape of this component is not well-constrained and can also be fit with a blackbody with $T \gtrsim 20000$), as well as the sum of those three models (blue line). The large majority of positive outliers shown in black are emission lines, not noise.}
    \label{fig:broadbandfit}
\end{figure*}

We use Molecfit \citep{kausch15, smette15} to correct the extracted VIS and NIR arm spectra for telluric absorption. In all bands, the flux is a few percent (4.5, 3.5, and 9\% in the NIR, VIS and UVB arms, respectively) higher in the first epoch, most likely due to slit losses in the second epoch with slightly worse seeing. We combine the two epochs by simple weighted averaging after applying the barycentric velocity correction and scaling the fluxes of the second epoch up to match those of the first epoch. The full spectrum is shown in Fig. \ref{fig:broadbandfit}.

\section{Analysis and results}\label{sec:res}
\subsection{The donor star}
\begin{figure}
    \centering
    \includegraphics[width=0.47\textwidth]{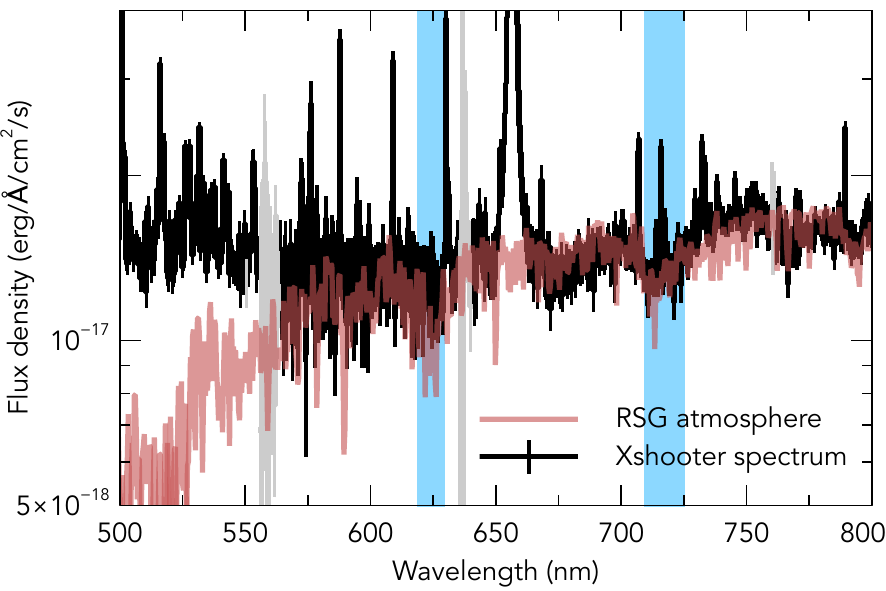}
    \caption{\xshoo data (black, with errorbars) in the 500-800 nm region containing TiO absorption features (indicated by the blue shaded regions; only the feature at 720 nm is visible in the data). Data points with low S/N are plotted in grey. The best matching MARCS model atmosphere is overplotted in red. Bluewards of 650 nm the data start to deviate from the model atmosphere --- an extra model component is necessary to fit this excess blue emission.}
    \label{fig:optical}
\end{figure}

The NIR part of the spectrum, redwards of $\sim 700$ nm, is dominated by absorption features that are typical of cool stars. Most prominent are the CO bandheads in the $H$- and $K$-bands and many molecular and atomic lines in the $J$- and $H$-bands. Meanwhile, in the optical \edit1{one of the characteristic TiO absorption bands is visible (see Figure \ref{fig:optical}).} These spectral features all point to the companion being a luminous, cool star \citep[e.g.][]{meyer98}. 

\begin{figure*}
    \centering
    \includegraphics[width=\textwidth]{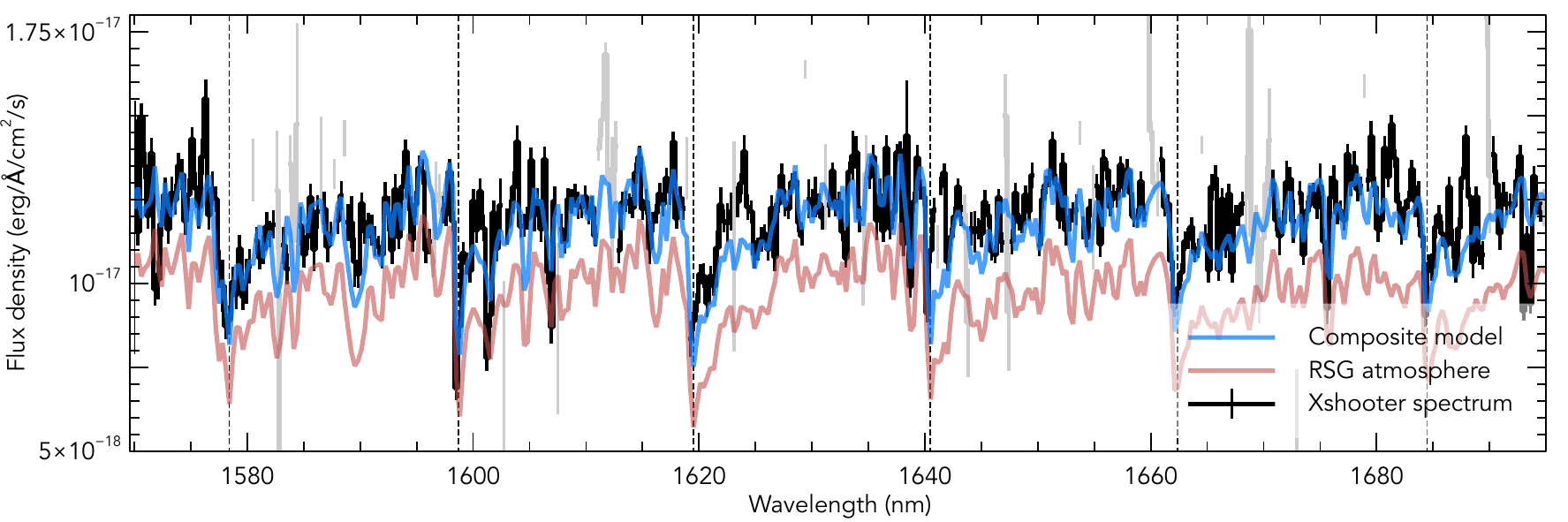}
    \caption{\xshoo data (black, with errorbars) in the 1575-1700 nm region containing several strong absorption features, most prominent of which are the CO bandheads (indicated with black dashed lines). Data points that are affected by strong telluric emission lines are plotted in grey. Overplotted are the best matching MARCS model atmosphere (red) and the composite model including an 1100 K blackbody to account for excess emission (blue).}
    \label{fig:abslines}
\end{figure*}

To obtain a temperature for the companion, we concentrate on the $H$-band region of the spectrum. The TiO features in the optical are well-known to be poorly reproduced by 1-D model atmospheres \citep{davies13}, while the $K$-band emission from \snulx is contaminated by dust emission which makes the stellar continuum difficult to place. The features we concentrate on in the $H$-band are the CO bandheads, which are known to be sensitive to effective temperature \citep[e.g.][see Figure \ref{fig:abslines}]{davies09}. 

The model grid we employ consists of synthetic spectra computed from MARCS model atmospheres \citep{MARCS} using TURBOSPECTRUM \citep{TURBOSPEC}. We investigate effective temperatures \Teff in the range 3400-4400 K, and gravities \logg between 0.0 and +1.0. For metallicity [Z], we compare the location of ULX-1 to the radial metallicity gradient as measured by \citet{Gazak15}, which indicates a value of [Z]$\simeq -0.1$. Therefore, we explore metallicities between [Z]=0.0 to -0.25. Finally, we search microturbulent velocities $\xi$ between 2 and 5 \vel, noting that the typical value for RSGs seems to be around 4 \vel \citep{davies15}. 

Our methodology is to fix \logg, $\xi$ and [Z], and find the best fit \Teff using $\chi^2$-minimization comparing the flux levels around the CO bandheads. We then repeat for all combinations of \logg, $\xi$ and [Z] within our permitted parameter space to find the full possible range of \Teff values. Irrespective of the combinations of \logg, $\xi$ and [Z], we consistently find temperatures within the range $3650 < T_{\rm eff} / K < 3900$. 

To find the luminosity $L_{\rm bol}$ of the companion, we assume that the RSG dominates the flux in the $H$-band (this is justified by our broad-band fit, see Figure \ref{fig:broadbandfit}). We then use the $H-K$ colours of the best-fitting models to estimate the $K$-band flux of the star. Since the bolometric correction (BC) at $K$ is roughly the same for all RSGs \citep[BC$_K\simeq$3, e.g.  ][]{davies13,davies-crowther-beasor18}, this then allows us to estimate a luminosity for the companion of $\log(L_{\rm bol}/L_{\odot}) = 4.25\pm0.10$. Converting the effective temperature and luminosity to an equivalent radius $R$, we find $R/R_{\odot} = 310 \pm 70$. 

We determine the radial velocity of the RSG through a cross-correlation of the \xshoo spectrum and the best-matching model atmosphere, using small regions around the strongest absorption lines in the 1150-1350 nm range. We find a radial velocity of $106 \pm 8$ \vel \citep[not corrected for the bulk motion of NGC~300 of $\sim 144$ \vel;][]{westmeier11}.

Two additional components are necessary to fit the full \xshoo spectrum (see Fig.~\ref{fig:broadbandfit}). Bluewards of 700 nm the continuum emission is dominated by an additional blue component, and there is also excess emission redwards of 1700 nm. The blue component can be represented by either a hot blackbody (T $\gtrsim 20,000$ K) or a power law with \edit1{spectral} index $\alpha \approx 4$; the S/N of the blue end of the \xshoo spectrum is too low to discriminate between these two options. The additional red component is well fitted by a blackbody of $\sim 1100$ K.  

\subsection{Emission lines}
The optical spectrum of \snulx is dominated by many emission lines that are not connected to the donor star. A full analysis of these lines will be presented in a separate paper; we focus here on the H$\alpha$ and He II $\lambda4686$ lines that were also analyzed by \citet{binder18}. All velocities mentioned below are relative to the rest wavelength in air of the respective lines. We did not correct for the bulk motion of NGC~300. 

\subsubsection{H$\alpha$ line}
\begin{figure}
    \centering
    \includegraphics[width=0.47\textwidth]{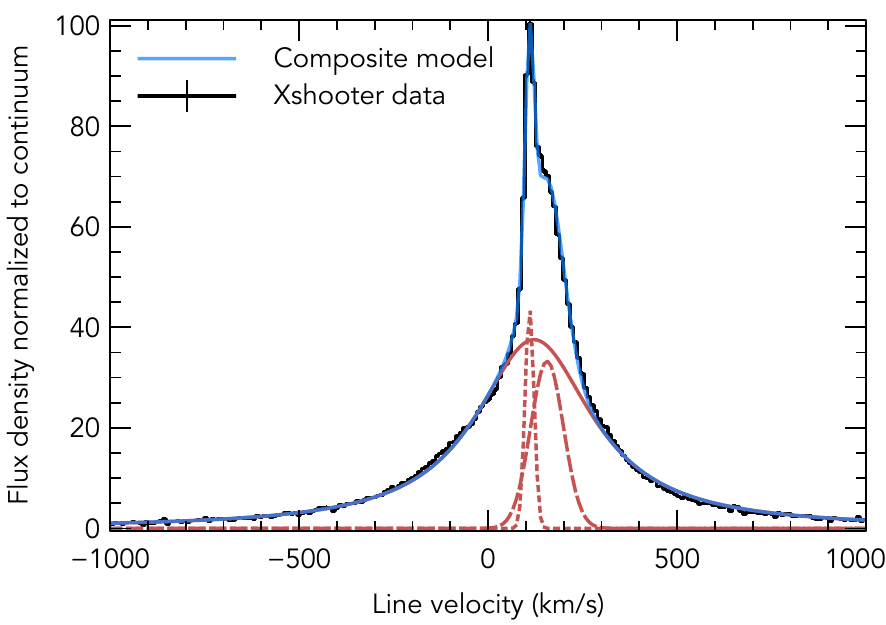}
    \caption{The H$\alpha$ line normalized to the continuum flux and shifted to a continuum of zero (black line with error bars). The zeropoint of the line velocity is at 656.276 nm. The three model components are plotted in red (narrow Gaussian, dotted line; redshifted, broader Gaussian, dashed line; broad Lorentzian, solid line) and the total model is plotted in blue.}
    \label{fig:halpha}
\end{figure}

The H$\alpha$ line in the spectrum of \snulx is best fit by three components: a narrow Gaussian with a FWHM of $28.1 \pm 0.5$ \vel, centered at $109.54 \pm 0.17$ \vel ($656.5158 \pm 0.0003$ nm); a broad Lorentzian with a FWHM of \edit1{$376 \pm 3$} \vel, centered at $121.4 \pm 0.6$ \vel ($656.5396 \pm 0.0014$ nm); and a second, \edit1{intermediate width,} redshifted Gaussian with a FWHM of $102.3 \pm 1.5$ \vel, centered at $155.9 \pm 0.7$ \vel ($656.6160 \pm 0.0015$ nm; see Fig.~\ref{fig:halpha}). The total \edit1{observed} flux in the line is $(8.76 \pm 0.08) \times 10^{-16}$ \flx.  \edit1{The equivalent width of the H$\alpha$ line is $-69.47 \pm 0.03$ \AA. The spectral resolution in the VIS arm corresponds to a velocity resolution of 26 \vel. The intrinsic FWHM is $10.6 \pm 0.5$ \vel for the narrow Gaussian component, $375 \pm 3$ \vel for the broad Lorentzian, and $99.0 \pm 1.5$ \vel for the intermediate width Gaussian. We fit the same three-component model to the individual epochs to check if shifts in the line velocities are a cause of broadening, but we do not find any variability in the central wavelenghts and FWHM of the three line components.}

\subsubsection{He II line}
\begin{figure}
    \centering
    \includegraphics[width=0.47\textwidth]{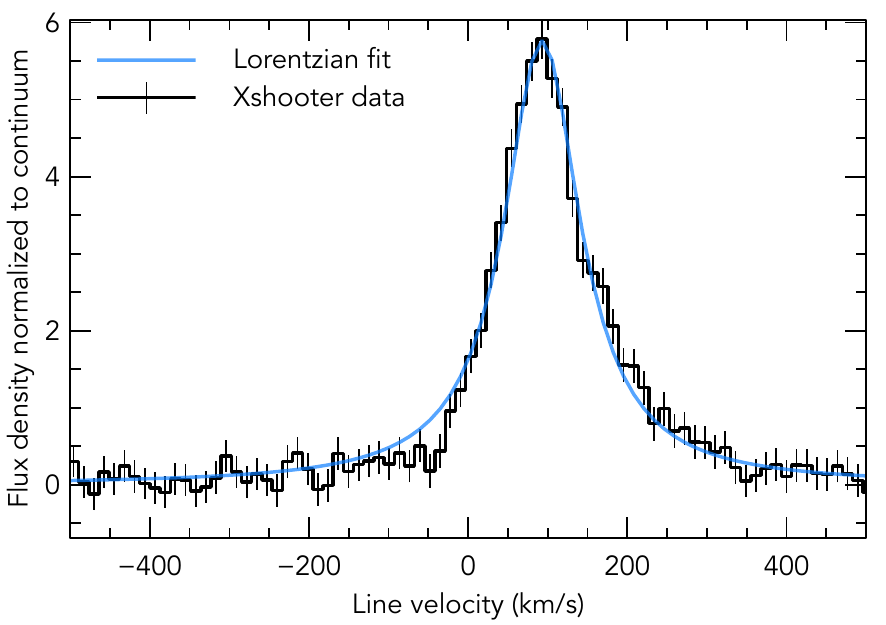}
    \caption{The He II line normalized to the continuum flux and shifted to a continuum of zero (black line with error bars). The zeropoint of the line velocity is at 468.575 nm. The Lorentzian fit to the line is plotted in blue.}
    \label{fig:he2}
\end{figure}

The He II line at 4686 \AA{} is best fit by a Lorentzian with a FWHM of $116 \pm 4$ \vel, centered at $93.0 \pm 1.3$ \vel ($468.720 \pm 0.002$ nm; see Fig.~\ref{fig:he2}). The \edit1{observed} flux in the line is $(2.40 \pm 0.08) \times 10^{-17}$ \flx. \edit1{The equivalent width of the line is $-1.63 \pm 0.06$ \AA. The spectral resolution in the VIS arm corresponds to a velocity resolution of 45 \vel. The intrinsic FWHM of the He II line is $107 \pm 4$ \vel.}

\section{Discussion and conclusions}\label{sec:disc}
\subsection{The nature of the donor star}
The shape of the continuum in the 750-1700 nm region of our \xshoo spectrum, combined with the properties of the absorption bands due to TiO and CO and the many absorption lines due to neutral metals, allow us to determine the properties of the donor star in \snulx for the first time. We find an effective temperature of 3650-3900 K and a bolometric luminosity $\log(L/L_\odot) = 4.25 \pm 0.1$, which implies an RSG with a radius of $310 \pm 70$ \rsun. The radial velocity of the star is $106 \pm 8$ \vel, consistent with the RSG being located in NGC~300 \citep[which has a radial velocity of 144 \vel and a maximum rotational velocity of 99 \vel][]{westmeier11}.

\begin{figure}
    \centering
    \includegraphics[width=0.47\textwidth]{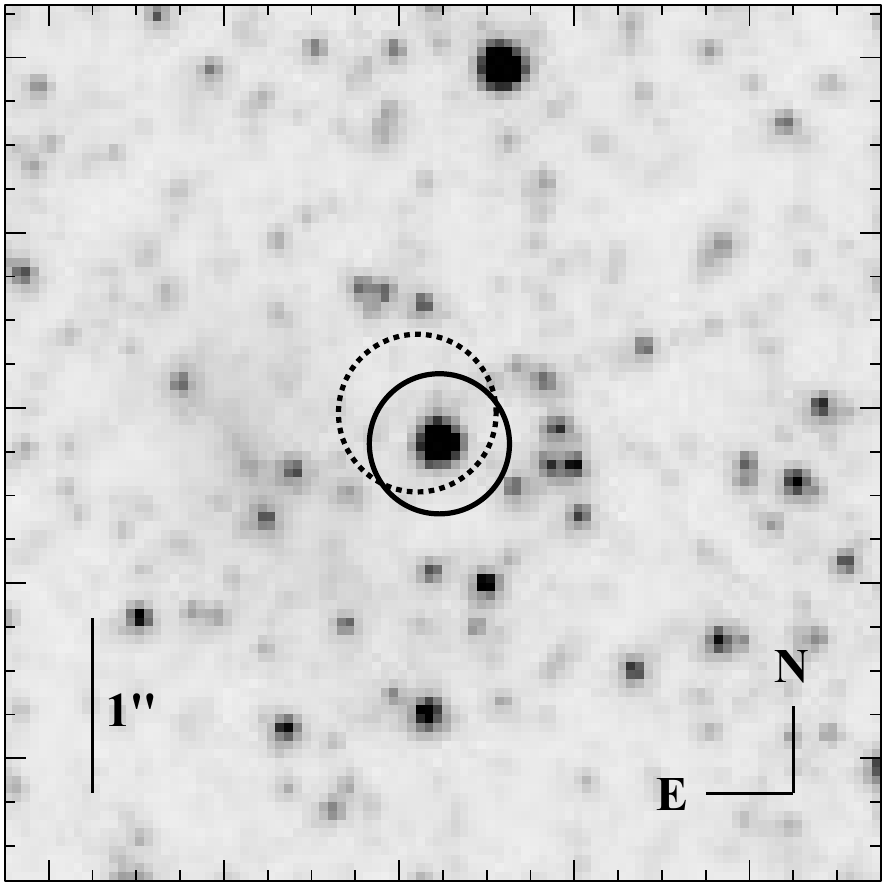}
    \caption{\hst F814W image taken post-outburst in July 2014 with the location of the X-ray source from  \chan \citep[$0.47''$ radius, dotted error circle,][]{binder16}. The solid circle centered on the optical source has a diameter of $0.8''$, the size of the widest \xshoo slit that we used, showing that there are no other sources contributing significantly to the optical/NIR emission.}
    \label{fig:hst}
\end{figure}

The connection between this RSG and the ULX is very secure. The position of the system that underwent the outburst in 2010 is accurately known \citep{monard10}. Non-detections of the progenitor in optical bands and consistent detections of the source in the optical post-outburst, in combination with its mid-IR behavior, indicate a dust-enshrouded progenitor system that became visible at optical wavelengths after most of the dust was destroyed in the 2010 outburst \citep{lau16,villar16}. \hst observations show the post-outburst optical source to be fairly isolated, and it is by far the brightest source in a $0.5''$ radius --- there is no other source nearby that could have contaminated our \xshoo spectra (see Figure \ref{fig:hst}). In addition, \citet{binder16} showed that the counterpart of SN2010da is the only optical source within the localization error of the ULX. The presence of emission lines, in particular the He II $\lambda4686$ line that is excited by X-ray photons, is additional evidence that the RSG and X-ray source belong to the same system.

The identification of the donor star as an RSG has important implications both for the current state of the system and for the interpretation of the 2010 SN impostor event. Previous donor star identifications, all based on photometric observations in different bands, have included an LBV \citep{binder16}, an sgB[e] star \citep{lau16} and a yellow supergiant \citep[][who did point out that the SED was consistent with an RSG as well but preferred a yellow supergiant based on the radius of the blackbody in the 2010 outburst]{villar16}. The presence of an RSG necessitates a fairly wide binary system, and the large donor-to-compact object mass ratio --- given that typical RSG masses are $\geq 8$ \msun --- means that stable Roche lobe overflow may not be possible (\citealt{vandenheuvel17}, but see also \citealt{quast19}). The slow wind speeds in RSGs (typically $20-30$ \vel, \citealt{richards98,vanloon01}) make \snulx a good candidate for focused wind accretion in a near-Roche lobe overflow system, as suggested by, e.g., \citet{elmellah19}.


\subsection{Orbital parameters}
The orbital period of \snulx is unknown. Contrary to other ULX pulsars, whose orbital periods have been determined or constrained from Doppler shifts of the pulse period \citep{bachetti14,israel17a,fuerst18,sathyaprakash19,rodriguezcastillo19}, the pulse period evolution of \snulx shows no sign of orbital modulations \citep{ray18,vasilopoulos18}. This is due in part to the very high absolute spin-up rate, which makes it difficult to detect small modulations to the spin period due to orbital motions. 

\edit1{The effective temperature and luminosity of the donor star are consistent with MESA evolution tracks of 8-10 \msun single stars \citep{choi16}. We use this as our best estimate for the mass of the donor star, with the caveat that this star evolved in an interacting binary system and, moreover, experienced a large outburst just a few years ago. A detailed investigation of its evolutionary history is necessary to truly determine its mass, but is beyond the scope of this letter.} 

Assuming an RSG with a mass of 8-10 \msun and a radius of $310 \pm 70$ \rsun, and a neutron star of 1.4-2.0 \msun, we can calculate the orbital period at which the RSG would fill its Roche lobe \citep{eggleton83}. For these parameters, the Roche-lobe filling orbital period would be 0.8-2.1 years, consistent with the constraints from X-ray timing measurements of the pulse period \citep[orbital period $\geq 1$ year,][]{ray18,vasilopoulos18} and with the possible mid-IR period of $\sim 400$ days suggested by \citet{lau16}. The orbital velocity of the RSG would be 8-10 \vel, and the orbital velocity of the neutron star 40-62 \vel. \edit1{If the donor star is more massive, the orbital period would be even longer and the orbital velocities lower.} The radial velocities of the two components are given by $v_\mathrm{orb}\sin(i)$; unless we are viewing the system at very low inclination, it is extremely unlikely that we will be able to detect radial velocity shifts for the RSG --- particularly since convective motions in RSG atmospheres can produce radial velocity shifts of similar magnitude \citep{ohnaka17}, which would make any radial velocity curve of the RSG unreliable unless the orbital period is confirmed independently. The radial velocity of the neutron star is likely high enough to be measured unless the system is seen at a very unfavorable inclination, either through pulsar timing if the absolute spin-up decreases sufficiently such that the orbital modulations become detectable, or through radial velocity shifts of emission lines that are formed in the accretion disc and co-moving with the neutron star. However, this measurement will be extremely challenging due to the long orbital period of the system, requiring many years of monitoring; the emission line radial velocity method specifically has been attempted before for other ULXs but radial velocity variations were found to be essentially random \citep{roberts11}. 

These numbers are in fact lower (upper) limits for the period (orbital velocities), as we do not know whether or not mass accretion occurs through Roche-lobe overflow in \snulx. If the neutron star is instead accreting material from the wind of the RSG (as suggested by, e.g., \citealt{elmellah19}) the orbital period may be longer and the radial velocities lower.

\subsection{Additional broadband components}
The broadband spectrum can only be fit by adding two extra components, a blue component that dominates the emission bluewards of $\sim 700$ nm and a red component that dominates the $K$-band emission. The excess red emission is also observed in \spitzer observations \citep{lau16, lau19} and is consistent with being due to warm dust with a temperature of $\sim 1100$ K. The presence of excess blue emission is consistent with predictions from theory \citep[e.g.,][]{copperwheat05, ambrosi18}, and is likely due to an irradiated accretion disc and/or irradiation of the donor star by the bright X-ray source. The strong variability of \snulx in X-rays indicates that this component, if it is indeed due to reprocessed X-ray emission, may be variable as well. Our \xshoo spectra were taken during a decrease of the X-ray flux of \snulx (see, e.g., the \swift light curve in \citealt{vasilopoulos19}); future observations taken at different X-ray flux levels are necessary to test this hypothesis. 

\subsection{Emission line structure}
The structure of the Balmer emission lines is best described with three components: a narrow Gaussian component, a broad Lorentzian that is redshifted by $\sim 15$ \vel with respect to the narrow Gaussian, and an intermediate width second Gaussian redshifted $\sim 45$ \vel. This structure, with a narrow, broad, and redshifted intermediate component, strongly resembles the line structure observed two weeks after the 2010 outburst by \citet{villar16}, \edit1{although both the intermediate Gaussian and broad Lorentzian components have become considerably narrower; their FWHM decreased from $\sim 500$ \vel to $\sim 100$ \vel and from $\sim 1000$ \vel to $\sim 375$ \vel, respectively.} The intermediate Gaussian component has also become less redshifted with respect to the narrow component, from $\sim 140$ \vel in 2010 to $\sim 45$ \vel now. \citet{villar16} modeled their late-time spectra as single Gaussians with a FWHM ranging from 300-600 \vel and an extended red wing, but these spectra were taken at lower resolution which likely precluded the detection of, for example, the additional narrow component. 

\citet{villar16} suggested that the narrow component seen in their early spectra was due to a pre-existing wind from an earlier RSG phase, while the intermediate \edit1{width} emission was powered by wind mass loss from the (blue or yellow supergiant) star post-outburst. Our identification of the post-outburst star as an RSG supports the claim that the narrow component of the Balmer emission lines is due to an RSG-driven wind; its FWHM of 10 \vel matches typical RSG wind speeds \citep{richards98,vanloon01}. This interpretation is also supported by the fact that the radial velocity of this component matches the radial velocity of the star itself. \edit1{Alternatively, the narrow component could be due to a surrounding photo-ionized nebula} as seen in other ULXs \citep[e.g.][]{pakull02,moon11}. 

The intermediate width emission is too broad to be due to a persistent stellar wind driven by the RSG, \edit1{but its width is consistent with shock-ionized ULX bubbles \citep{pakull10}.  An ultrafast outflow (at 0.22 c) was detected in this system in X-rays by \citet{kosec18}, showing that strong winds are launched from the disc. Either the intermediate or broad H$\alpha$ component may be due to a slower and cooler disk wind \citep[similar broad hydrogen and sometimes helium emission lines have been detected in other ULXs and interpreted as evidence for a disk wind,][]{fabrika15}. One of these components may also be associated with material expelled by the star in the 2010 outburst.}

\subsection{He II line luminosity}
The luminosity in the He II $\lambda 4686$ line can be used to infer the ionizing flux in the 54-200 eV range and, in combination with the measured X-ray flux in that range, to determine the beaming factor of the soft X-rays, as the He II emission is isotropic \citep{pakull86}. \citet{binder18} performed this calculation for \snulx using their Gemini spectra and simultaneous \swift observations obtained in June and July 2017, when the source was in the ULX regime; they found that the number of He-ionizing photons calculated from the He II luminosity and from the X-ray luminosity (calculated from the X-ray flux assuming isotropic emission) were roughly equal, within a factor of 2. This indicates that in the super-Eddington state, the soft X-rays (below 0.2 keV) were not strongly beamed, in contrast with the higher energy X-rays that must be more strongly beamed given the high pulsed fraction in \snulx \citep[up to $\sim 76\%$ above 2 keV,][]{carpano18}. It also shows that the He II emitting gas surrounds the ULX with a covering factor close to one. 

\edit1{\snulx was monitored regularly in the X-rays with \swift in 2017 and 2018. \citet{vasilopoulos19} show that the X-ray luminosity was stable at a level of $\sim 5\times 10^{39}$ \lum until February 2018, when it started to drop until it settled down at around the Eddington luminosity of a neutron star, a factor $\sim 25$ lower than its peak luminosity, in September 2018 (their figure 1). It was still at this lower X-ray flux at the time our \xshoo spectra were taken in October. \nicer and \swift observations of the pulse period obtained in August and November 2018, when the source rebrightened in X-rays for a short time, show that the spin-up rate of the pulsar remained constant throughout 2018 (at $\dot \nu \approx 4 \times 10^{-10}$ s$^{-2}$) and did not vary with the X-ray flux, as would be expected if the flux decrease was due to a decrease in the mass accretion rate \citep{ray18,vasilopoulos19}. In addition, the ratio of $0.3-1.5$ keV to $1.5-10.0$ keV photons, as seen by \swift, did not change when the X-ray flux dropped, as would be expected if the source had undergone a state change or if the obscuring column density had increased.} \citet{vasilopoulos19} suggested that the observed lower X-ray flux was due to increased obscuration by an optically thick outflow from the radiation-dominated inner accretion disk that moved into our line of sight due to Lense-Thirring precession.

The flux we measure in the He II line in our October 2018 spectrum corresponds to a luminosity of \edit1{$1.19 \pm 0.04 \times 10^{34}$ \lum, after correcting the line flux for foreground Galactic extinction \citep[$E(B-V) = 0.011$;][]{schlafly11}} and assuming a distance of 2.0 Mpc to NGC~300. This is a factor of $\sim 70$ lower than the luminosity reported by \citet{binder18}, and the ionizing photon flux in the 54-200 eV range inferred from this line luminosity is within a factor of two of the photon flux \edit1{inferred from the X-ray flux at the time of our observations,} assuming the spectral shape has indeed remained constant. This drop in the He II luminosity indicates that the {\it total} X-ray luminosity must have decreased, rather than merely that in our line of sight. This argues against partial obscuration by a precessing accretion disk, which would not affect the total X-ray luminosity seen by the He II emitting region. Instead, it suggests either an overall increase of obscuring material, possibly due to increased outflow from the accretion disk, or an intrinsic drop in the X-ray luminosity. However, it is unclear how either of these scenarios can be made compatible with the lack of variability in the X-ray spectral shape and, for the latter scenario in particular, with the constant spin-up rate of the pulsar.

\subsection{Conclusions}
\snulx is the second known ULX pulsar with a supergiant donor star (the other being \pth, which has a B9I companion, \citealt{motch11,motch14}). The other four ULX pulsars known to date all have orbital periods of only a few days \citep{bachetti14, israel17a, sathyaprakash19, rodriguezcastillo19}, implying that they likely have main sequence or Hertzsprung gap donors, although those are too faint to be detected in these extragalactic systems \citep{heida19}. Among the general population of ULXs, three other sources have counterparts that have been spectroscopically identified as RSGs \citep{heida15a, heida16}, and \citet{lau19} identified two others as likely RSGs based on their mid-IR colors. 

This small but significant fraction (two out of six ULX pulsars, and at least 4-6 out of $\sim 150$ ULXs within 10 Mpc) with supergiant donors is an extremely interesting subset of ULXs, especially since they may be progenitors of binary compact object mergers \citep[e.g.][]{belczynski16}. \edit1{ULXs with supergiant donors} are currently underproduced by binary population synthesis codes \citep[e.g.][who find RSG donors in $<1\%$ of systems]{wiktorowicz17}. This may be due to the fact that these codes don't take mass transfer through (focused) wind mass loss into account, and shows that this phenomenon may play an important role in ULXs with (red) supergiant donor stars.

SN2010da/\snulx is the only ULX to date that is also associated with an ILOT; vice versa, it is also the only ILOT that has turned into an X-ray binary post-explosion, as well as the only one that contains a spectroscopically confirmed red supergiant star. This unique system has the potential to offer insights into the formation of supergiant ULXs and episodic mass loss from evolved massive stars; \edit1{for example, it is possible that the 2010 outburst was related to a common envelope phase, or to a tidal interaction in a very eccentric binary system. Long-term monitoring of both the X-ray source and the optical/NIR spectrum is necessary to determine the orbital period and discriminate between different scenarios.} Late-time follow-up of similar ILOTs with X-ray telescopes may reveal other supergiant ULXs and shed light on the evolutionary pathways that produce these systems.

\acknowledgments
MH would like to thank M. Bachetti, E. Levesque, G. Vasilopoulos, I. el Mellah, S. Guillot and E. Quatart for useful discussions. Based on observations collected at the European Organisation for Astronomical Research in the Southern Hemisphere under ESO programme(s) 102.D-0535(A).
\vspace{5mm}
\facilities{VLT(\xshoo)}

\software{Astropy \citep{2013A&A...558A..33A}, Reflex \citep{reflex}, Molecfit \citep{kausch15, smette15}, TURBOSPECTRUM \citep{TURBOSPEC}
}

\bibliography{bibliography}



\end{document}